\documentclass[a4paper,11pt]{article}
\usepackage{pos}

\usepackage[numbers]{natbib}
\title{Observational constraints on the blazar jet wobbling timescales}
 \ShortTitle{Observational constraints on the blazar jet wobbling timescales}

\author[a]{J.~Jury\v{s}ek}
\author[a]{V.~Sliusar}
\author[]{D.~Moulin}
\author[a]{R.~Walter}

\affiliation[a]{\textit{D\'epartement d'Astronomie, Universit\'e de Gen\`eve, Chemin d'Ecogia 16, CH-1290 Versoix, Switzerland}}

\emailAdd{jakub.jurysek@unige.ch}

\abstract{Blazars are a subclass of radio-loud active galactic nuclei (AGNs), where the jet is aligned close to the line of sight. Blazars emission is dominated by non-thermal processes, where Doppler boosted radiation originates from a relativistic population of charged particles within the jet. From radio to TeV energies, blazars are highly variable on timescales from minutes to several months. There are several mechanisms proposed to explain variability, including changes in the viewing angle of the jet, propagating along the rotation axis of the accretion disc. The misalignment of a supermassive black hole (SMBH) spin and the angular momentum of the accretion disc yields to Lense-Thirring precession of such tilted disc, which leads to the variation of Doppler beaming. Such scenario is supported by radio observations of jet precession observed in some AGNs. The radio-emitting regions, however, are located far from the central engine, and thus the observed time scales in this band can be affected by e.g. a variation of the bulk Lorentz factor along the jet.

In this contribution, we derive expected time scales of the jet wobbling using SMBH masses and compare them with the time intervals between flares in long-term (over $\sim15$ years) X-ray light curves of bright blazars observed by {\it Swift}-XRT. We found that for Mrk 421, Mrk 501 and 3C 273, the derived time scales are consistent with the observational constraints, while for 1ES 1959+650 we are mostly limited by uncertainty in the Doppler beaming factor.}

\FullConference{37$^{\rm{th}}$ International Cosmic Ray Conference (ICRC 2021)\\
		July 12th -- 23rd, 2021\\
		Online -- Berlin, Germany}

\begin{document}
\maketitle

\section{Introduction}

Blazars are a subclass of AGNs, characterized by relativistic jets pointing towards the observer. Emission of blazars is dominated by the jet in a wide range of wavelenghts from radio to very high energy gamma-rays. Blazars can be sub-divided to Flat Spectrum Radio Quasars (FSRQs) showing broad emission lines, and BL Lacs with no or only weak emission lines in their spectra. The spectral energy distribution (SED) of blazars can be characterized by two humps. The low energy hump is commonly believed to originate from the sychrotron emission of relativistic leptons in the jet, while the mechanism behind emission forming the high energy hump is still debated \cite[e.g.][]{2016CRPhy..17..594D}. Blazars are also characterized by a rapid variability at all wavelengths on timescales from minutes to months. The mechanism behind such variability is still not clear even though several were proposed, including changes of the viewing angle of the emitting region in a twisted or wobbling jet \cite[e.g.][]{2017Natur.552..374R}.

Detailed General Relativity magnetohydrodynamic simulations of a tilted disc-jet systems show that the disc precess together with the jet around the spin vector of the SMBH. During this process the tilt angle is decreasing, and eventually aligns with the SMBH spin on the accretion timescale \cite{2018MNRAS.474L..81L}. In addition to the precession, the simulations showed disc-jet wobbling by several degrees in amplitude on relatively short timescale of about $10^3 - 10^4 t_\mathrm{g}$ in the source frame, where $t_\mathrm{g} = r_\mathrm{g} / c$. As suggested by Liska et al. \cite{2018MNRAS.474L..81L}, such variations of the jet viewing-angle could boost jet emission in and out of the line-of-sight, resulting in observations of high-energy flares on a timescale of the wobbling.

In this contribution, the goal is to constrain a typical intervals between the flares for selected bright blazars and compare them with the expected wobbling timescales dependent on the SMBH mass $M_\mathrm{SMBH}$. The time intervals between flares in the units of $t_\mathrm{g}$ transformed in the source frame can be expressed as
\begin{equation}
\frac{\Delta \tau}{t_\mathrm{g}} = k \frac{\delta \Delta t}{(1 + z) M_\mathrm{SMBH}}, \; k = 8.7 \times 10^{9} \, \mathrm{M_\odot \, day^{-1}},
\label{eq.dtau}
\end{equation}
where $\delta$ is the Doppler beaming, $z$ is the redshift, $\Delta t$ is the time interval between flares in the observer's frame. While $z$ for the bright blazars in our sample are determined with high precision, uncertainties of the other parameters cannot be neglected. In the following sections, we will carefully evaluate possible ranges of $\delta$ and $M_\mathrm{SMBH}$, determined by various methods for individual sources, and assess whether they are narrow-enough to confirm or reject the hypothesis of jet-wobbling induced flares.

\subsection{SMBH masses}\label{sec.masses}


According to the unification scheme of AGNs, the Broad Line Region (BLR) is assumed to extent not farther than a few hundreds of Schwarzschild radii from the SMBH in the center. If the motion of individual clouds in the region is dominated by the gravitational force, the $M_\mathrm{SMBH}$ can be estimated from the virial equation $M_\mathrm{SMBH} = f R_\mathrm{BLR} v^2 G^{-1}$, where $R_\mathrm{BLR}$ is the radius of the BLR, $v$ is the mean velocity of the clouds, $G$ is gravitational constant, and $f$ is a factor taking into account geometry and kinematics of the region \cite[e.g.][]{2002MNRAS.331..795M}. The velocity $v$ can be obtained directly from FWHM of broad emission lines, and if a long-term spectroscopy of a source is available, the distance of the emitting region from the center $R_\mathrm{BLR}$ can be determined from the time lag between emission in lines and the continuum (the so called reverberation method \cite{Peterson_1993}). It also turned out that there is a correlation between $R_\mathrm{BLR}$ and optical monochromatic continuum luminosity $L_\mathrm{5100}$ \cite{2000ApJ...533..631K}, or $\mathrm{H_\beta}$ emission line luminosity \cite{2011JApA...32..209W}, which is less affected by the luminosity of the jet and therefore provides less biased values of $R_\mathrm{BLR}$.

For both quiescent and active galaxies, there is a relatively tight correlation between $M_\mathrm{SMBH}$ and bulge velocity dispersion $\sigma$ \cite{Ferrarese_2001}. Direct measurement of $\sigma$ for distant AGNs is difficult, but most of the AGN host galaxies are elliptical, and for those $\sigma$ can be determined from effective radius and average surface brightness using the so called fundamental plane relation \cite{2003ApJ...595..624F}. The results obtained using this method was adopted for all BL Lacs in our sample, where the virial method cannot be used due to absence of the emission lines in their spectra.

\subsection{Doppler beaming}
Blazar jet emission is experiencing the relativistic beaming effects, which increase the brightness and and compress timescales in the observer's frame. The observed $\Delta t$ between flares have to be therefore transformed to the source frame. The Doppler beaming factor $\delta$ is defined as $\delta = [\Gamma (1 - \beta \cos \Theta) ]^{-1}$, where $\Gamma = 1/\sqrt{1 - \beta^2}$ is the bulk Lorentz factor and $\Theta$ is the viewing angle of the jet.

For FSRQs, $\delta$ can be estimated from apparent speed $\beta_\mathrm{app}$ of radio blobs observed by VLBI \cite[e.g.][]{Jorstad_2005}. The highest detected $\beta_\mathrm{app}$ puts a lower limit on the bulk Lorentz factor $\Gamma \geq \sqrt{1 + \beta_\mathrm{app}^2}$. The viewing angle can be then taken as $\Theta \leq 1/\sin{(\beta_\mathrm{app}^{-1})}$ and thus $\delta \approx \beta_\mathrm{app}$. For the only FSRQ in our sample - 3C 273 - $\delta$ estimated from radio observations of the blobs is also consistent with the results of SED modeling, as shown later in Section~\ref{sec.results}.

For BL Lacs, on the other hand, the observed radio knots are stationary or even show inward motion \cite[e.g.][]{2019ApJ...874...43L}, while  $\delta$ of the order of 10 is usually needed in one zone Synchrotron Self-Compton (SSC) scenario to explain observed SED. Such discrepancy is called the bulk Lorentz factor or Doppler crisis and hasn't yet been sufficiently explained \cite[e.g.][]{2018mgm..conf.3074P}. In this analysis, we use values resulting from leptonic SED models only. 

As the Doppler beaming can be highly variable along the jet and can also vary in time (due to long term precession of the jet or even the jet wobbling itself), we tried, if possible, to consider only the values of $\delta$ resulting from SED modeling between flares in quiescent state of a source, or the values obtained from averaged SED over a long period of observation. The range of $\delta$ determined for individual sources is discussed in the Section~\ref{sec.results}.

\section{Distributions of time intervals between flares}

We selected 4 bright blazars from the Fermi 4LAC catalog \cite{Ajello_2020}, with well covered {\it Swift}-XRT light curves and reliably determined mass: Mrk 421, Mrk 501, 1ES 1959+650 and 3C 273. For the light curve extraction, we used the online tool\footnote{\url{https://www.swift.ac.uk/user_objects/}}, using hourly binned data, and combined both photon counting and window timing modes of observation of {\it Swift}-XRT. Most of the sources are, however, so bright that the observations are conducted mostly in the Window timing mode. In order to get better SNR, we used the data from both {\it Swift}-XRT channels combined in the energy range of 0.3 - 10 keV.

\begin{figure}[!t]
\centering
\begin{tabular}{c}
\includegraphics[width=.98\textwidth]{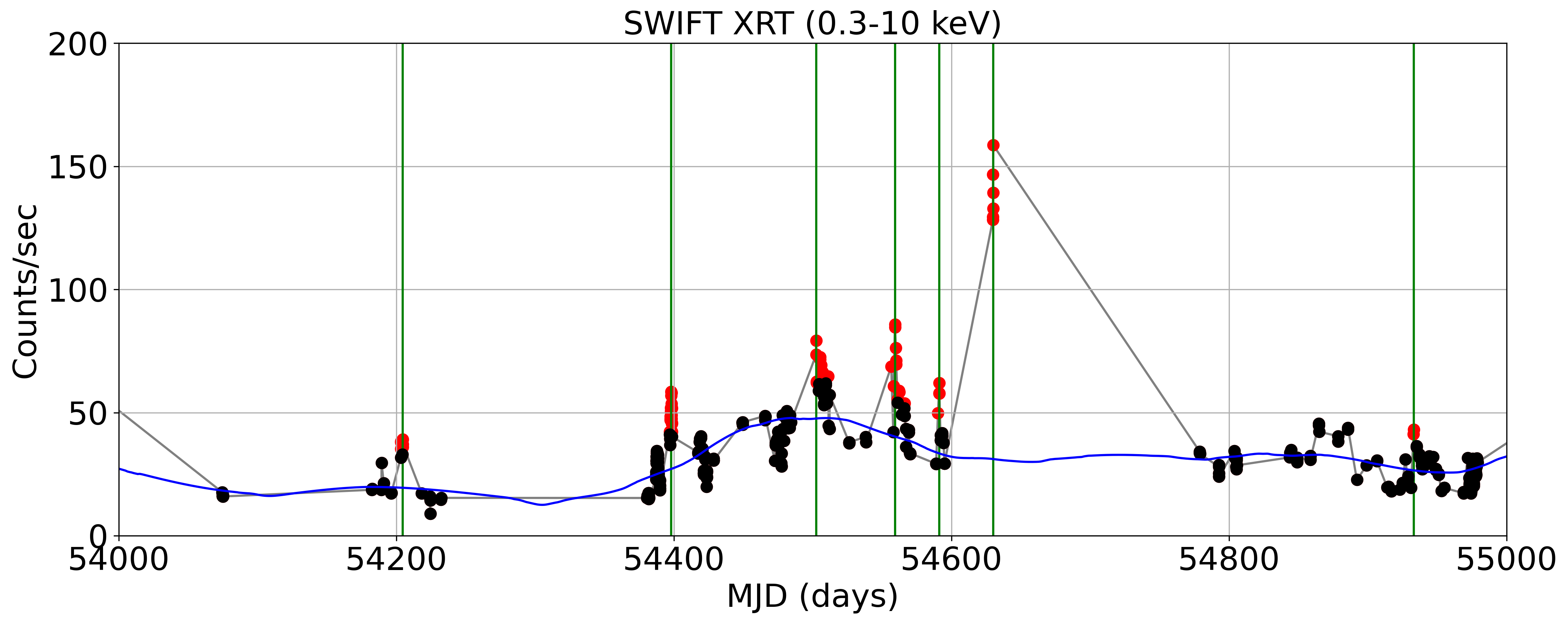}\\
\end{tabular}
\caption{Flares (marked by green lines) detected in a part of the {\it Swift}-XRT light curve of Mrk 421. The blue line shows the long term variability, and the red points are the suspected flares detected by our algorithm.}
\label{fig.swift_lc}
\end{figure}

To detect individual flares, we first applied second order Savitzky-Golay filter \cite{savitzky64} with the window length of about 200 days in order to subtract the long term variations in the light curves. For the purpose of the filtering, the data was re-sampled to get evenly spaced data set. High amplitude flares, however, can affect the filter significantly, and thus we run the filter iteratively. Usually $\sim4$ iterations were sufficient to subtract the long term variations so that the residuals were normally distributed. The flare candidates were then identified in the detrended data using $95\%$ percentile cut on the count rate distribution. After this fully automatic step, we also checked the detected flares visually and rejected false detections, or added the flares missed by the algorithm. For each flare detected, we demanded at least two data points significantly (>$2 \sigma$) above the slowly varying averaged light curve. An example of the {\it Swift}-XRT light curve of Mrk 421 with the flares detected is in Figure~\ref{fig.swift_lc}.

\begin{figure}[!t]
\centering
\begin{tabular}{cc}
\includegraphics[width=.45\textwidth]{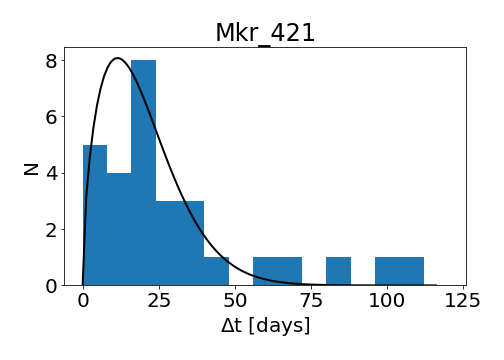} & \includegraphics[width=.45\textwidth]{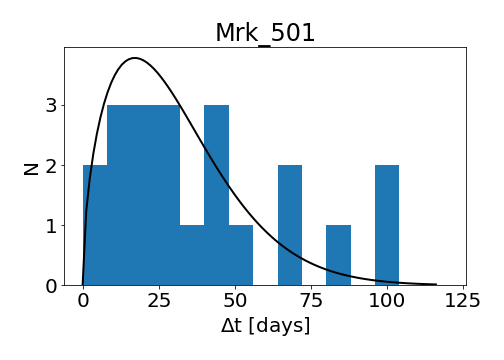} \\ \includegraphics[width=.45\textwidth]{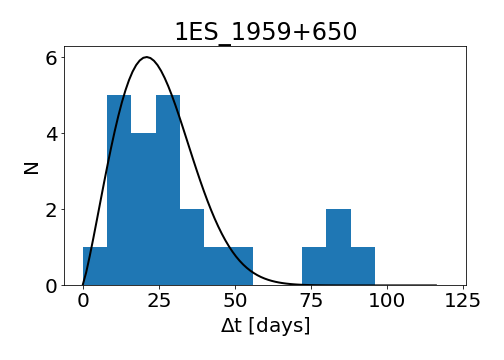} & \includegraphics[width=.45\textwidth]{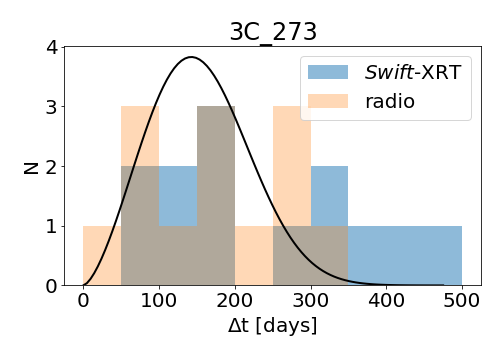} \\
\end{tabular}
\caption{Distributions of the time intervals between flares $\Delta t$, where the first peaks are approximated by the Weibull distribution (black lines).}
\label{fig.dt_obs}
\end{figure}

The distributions of time intervals between the flares in the observer's frame $\Delta t$ are shown in Figure~\ref{fig.dt_obs}. In order to constrain a possible range of the jet wobbling timescale, we assume that $\Delta t$ can be approximated by Weibull distribution, with suitable property of $\Delta t$ being always greater than 0. Fitting by Gaussian distribution, would lead to a significant leak of the $\Delta t < 0$, which is non-physical. For each distribution, we also performed Kolmogorov-Smirnov test to verify that the zero hypothesis (both distributions are the same) cannot be rejected. P-values of the KS-test was $> 0.85$ for all four sources.

We also assumed that the secondary peaks in the $\Delta t$ distribution are due to missing flares that were too weak to be detected by our algorithm, or fell in the periods not covered by the {\it Swift}-XRT observations in-between the flares detected. If that is the case, the longer time intervals would split in half and contribute to the first peak. Note that the distribution of $\Delta t$ for Mrk 421, peaking at $\Delta t \approx 20$ days, is consistent with the same distribution determined on the FACT light curve in the TeV range \cite{2021A&A...647A..88A}. 

For some FSRQs, the individual X-ray flares can be associated with ejection of radio knots, that travels as shock waves down the jet \cite[e.g.][]{2020MNRAS.492.3829L}. We therefore compared the time intervals between X-ray flares for 3C 273, with radio observations of superluminal knots ejection times, observed on a parsec scale by VLBA at 43 GHz \cite{2006A&A...446...71S, 2017ApJ...846...98J, 2020MNRAS.492.3829L}. Keeping in mind that the knots ejection rate varies in time \cite[e.g.][]{2012IJMPS...8..356J}, we considered only the knots ejected in the time range of the analyzed {\it Swift}-XRT light curves. The distribution of $\Delta t$ for 3C 273 is shown in Figure~\ref{fig.dt_obs} and one can note a double peak structure, similar to that present in the $\Delta t$ distribution for BL Lacs, where the distribution for radio knots is consistent with {\it Swift}-XRT.

\section{Notes on individual sources}\label{sec.results}

The SMBH masses of all BL Lacs in our sample were obtained combining results of different authors using $M_\mathrm{SMBH} - \sigma$ relation, discussed in Section~\ref{sec.masses}, taking into account uncertainties of measurements \cite{Wang_2004, 2008MNRAS.385..119W, 2003ApJ...595..624F, 2003ApJ...583..134B, Woo_2005, 2002ApJ...579..530W, 2002AA...389..742W}. For 3C 273,  we used the SMBH mass given by recent results of the GRAVITY collaboration \cite{2018Natur.563..657G}, who observed spatially resolved BLR. 

In this section, we discuss in detail the range of Doppler factors determined for individual sources, which are crucial for the determination of $\Delta \tau$. All the parameters are summarized in Table~\ref{tab.results}.

\subsection{Mrk 421}

Even though there is a degeneracy between $R, B$ and $\delta$, parameters and the size of the emission region was estimated with large uncertainty from minimum variability time scale varying from hours to days, Katarzynski et al. \cite{2003A&A...410..101K} were able to describe quiescent state of Mrk 421 with one-zone leptonic SSC model, using $\delta \approx 20$. Later in 2009, a 4.5 month long multiwavelength campaign was organized when Mrk 421 was mostly in low activity state, leading to a successful model of the averaged SED with $\delta = 21$ \cite{2011ApJ...736..131A}. The size of the emitting region was constrained from minimum variability time scale to be longer that 1 day during the campaign, which turned out to be consistent with the size of the VLBA core. Similarly Fraija et al. \cite{2017ApJS..232....7F} found in their long-term study study $\delta = 20$ for averaged SEDs of Mrk 421 in quiescent states between 2008 and 2012. Another analysis of long-term data resulted in estimation of $\delta$ in two quiescent periods as $\approx15$ and $\approx38$ \cite{Bartoli_2016}. Models of daily binned SEDs from 13 consecutive nights during flaring period of Mrk 421 show wide range of $\delta$ \cite{2016MNRAS.463.4481Z}, but comparing with the {\it Swift}-XRT light curve, we can select the nights of the lowest activity and constraint the range of $\delta \in (25, 35)$. 

\subsection{Mrk 501}
The analysis of data from a multiwavelenth campaign conducted during 2009 showed that he averaged SED can be described by a one-zone SSC model, with $\delta \in (11, 14)$, considering a limit of the emission zone radius derived from minimum variability time scale \cite{2011ApJ...727..129A}. That is consistent with the results obtained by Bartoli et al. \cite{Bartoli_2012}. Shukla et al. \cite{Shukla_2014} using multiwavelength data obtained in 2011 showed that the SED can be better described by two-component SSC, where both blobs are boosted by the same $\delta \approx 12$, which turned out to be stable for the whole period of the observations. 

\subsection{1ES 1959+650}
From two multivawelength flare light curves in 2016, $\delta$ was constrained by a leptonic one-zone SSC model \cite{2020A&A...638A..14M} as $\delta \in (30, 60)$. That is consistent with the results of other authors, deriving for high state $\delta = 20$ \cite{Krawczynski_2004}, or $40$ \cite{Aliu_2013}. Low state SED were also modeled in a few cases, giving $\delta = 25$ \cite{Aliu_2014} or $30$ \cite{Aliu_2013}. Provided that the most flares occurred in very active high state of the source, and that we have no other limits on $\delta$, we need to consider relatively wide interval of possible boosting $\delta \in (20, 60)$ for this source.

\subsection{3C 273}

3C 273 featuring an extended jet showing superluminal motion, and thus for this source, $\delta$ can be estimated from apparent speed $\beta_\mathrm{app}$ of radio blobs. Polarimetric measurements of the radio knots with VLBA lead to $\delta = 9 \pm 1.4$, which is also consistent with $\delta$ estimated from variability timescale of the superluminal jet components observed in radio \cite{Jorstad_2005}. Later observations of the radio knots on a parsec scale from the center showed $\beta_\mathrm{app}$ in a wider range from 5 to 15 \cite{2013AJ....146..120L}. Consistently with that, averaged multiwavelength SED was successfully modeled by one-zone leptonic SSC scenario, resulting in $\delta \approx 14$ \cite{DILTZ201463}. Note that the precession of the 3C 273 jet was observed with period of about 16 yrs, but the accompanying $\delta$ variations seem to fall in the range of $\beta_\mathrm{app}$ observed in radio \cite{2014MNRAS.437..489B}. 

\begin{table}
    \centering
    \begin{tabular}{r|cccc}
        Source & $z$ & $\delta$ & $M_\mathrm{BH}$ & $\Delta \tau / t_\mathrm{g}$ \\
        & & & $[10^8 \, M / M_\odot]$ & $[10^4]$ \\
\hline
		Mrk 421 & 0.0308 & 20 - 35 &	$1.8 \pm 0.3$  & 0.3 - 5.6  \\
        Mrk 501 & 0.034 & 11 - 14 & $5.7\pm3.0$ &  0.05 - 1.9 \\
		1ES 1959+650 & 0.048 & 20 - 60 & $1.2\pm0.4$ & 1.2 - 16 \\
		3C 273 & 0.158 & 5 - 15 & $2.6\pm1.1$ &  0.75 - 12.5 \\ 
\hline
    \end{tabular}
    \caption{Possible ranges of $\delta$ and $M_\mathrm{SMBH}$ and $95\%$ confidence intervals of $\Delta \tau / t_\mathrm{g}$.}
    \label{tab.results}
\end{table}

\begin{figure}[!t]
\centering
\begin{tabular}{cc}
\includegraphics[width=.45\textwidth]{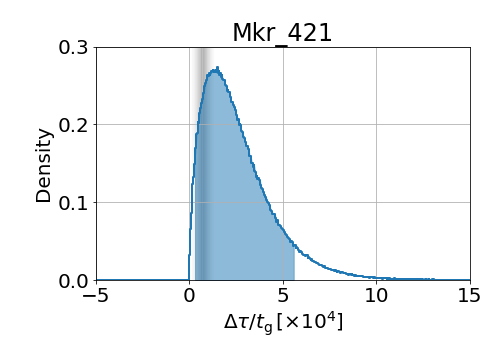} & \includegraphics[width=.45\textwidth]{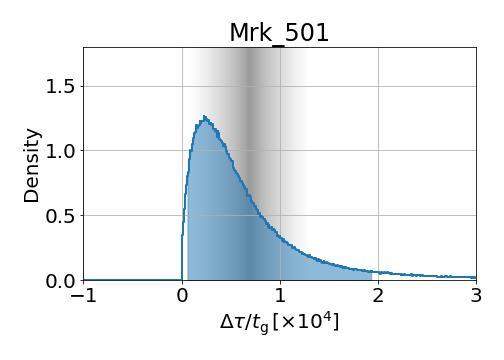} \\
\includegraphics[width=.45\textwidth]{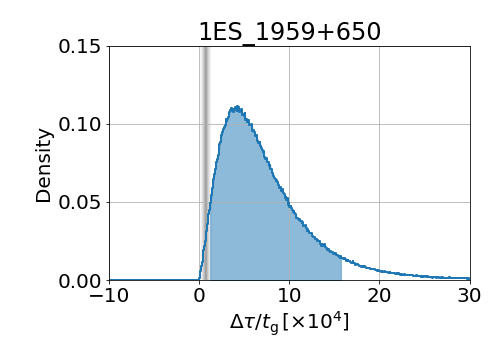} & \includegraphics[width=.45\textwidth]{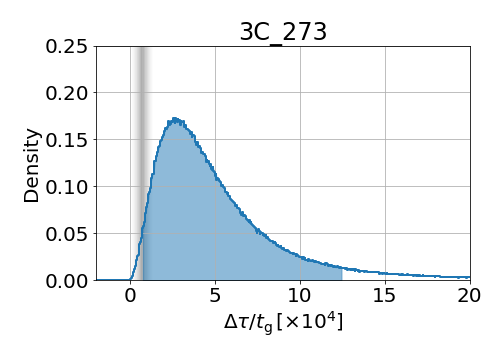} \\
\end{tabular}
\caption{Distribution (blue line) of estimated $\Delta \tau / t_\mathrm{g}$ for individual sources. Filled regions represent $95\%$ confidence intervals. The fuzzy grey regions show the expected range of jet wobbling timescale $10^3 - 10^4$ according to \cite{2018MNRAS.474L..81L}.}
\label{fig.dtau}
\end{figure}

\section{Conclusions}

In Figure~\ref{fig.dtau}, there are distributions of $\Delta \tau / t_\mathrm{g}$ for individual sources, obtained by Monte Carlo sampling of the Eq.~\ref{eq.dtau} within the observed ranges of $\delta$ and $M_\mathrm{SMBH}$. Observational constrains on $\Delta \tau / t_\mathrm{g}$ are rather weak, mostly due to the uncertainties of $\delta$, even if the results obtained from a single jet emission scenario were only considered.

Liska et al. \cite{2018MNRAS.474L..81L} showed that in the jet wobbling scenario, substantial changes of the viewing angle with amplitude of several degrees can occur on a timescale of about $\Delta \tau / t_\mathrm{g} \approx 10^3 - 10^4$, which is also shown in the Figure~\ref{fig.dtau} as fuzzy regions. For exact determination of the jet wobbling timescale, however, a detailed analysis of a power spectrum of the tilt angle variations simulated on a longer interval would be necessary. 

The jet wobbling timescale falls in the $95\%$ confidence interval of $\Delta \tau / t_\mathrm{g}$ for Mrk 421, Mrk 501 and 3C 273. For 1ES 1959+650, the timescale of blazar flares constrained by observations tends to be higher, even though the consistency with the jet wobbling scenario cannot be ruled out as the lower confidence limit is $\Delta \tau / t_\mathrm{g} \approx 1.2 \times 10^4$.

\bibliographystyle{JHEPe}
{\footnotesize
\bibliography{references}}

\end{document}